\documentclass[journal,twoside,web]{ieeecolor}
\usepackage{etoolbox}
\makeatletter
\@ifundefined{color@begingroup}%
  {\let\color@begingroup\relax
   \let\color@endgroup\relax}{}%
\def\fix@ieeecolor@hbox#1{%
  \hbox{\color@begingroup#1\color@endgroup}}
\patchcmd\@makecaption{\hbox}{\fix@ieeecolor@hbox}{}{\FAILED}
\patchcmd\@makecaption{\hbox}{\fix@ieeecolor@hbox}{}{\FAILED}
\usepackage{generic}
\usepackage{cite}
\usepackage{amsmath,amssymb,amsfonts}
\usepackage{algorithmic}
\usepackage{graphicx}
\usepackage{textcomp}
\begin{document}
\title{Candidate overtone shear horizontal SAW resonators in thin-film lithium niobate for intermodal acousto-optic modulation}
\author{Wenbing Jiang, Xuankai Xu, \IEEEmembership{Graduate Student Member, IEEE}, Yu Guo, Boyu Zhang, Lihui Jin, Xiao Chen, Tao Wu, \IEEEmembership{Senior Member, IEEE}, and Libing Zhou
\thanks{This work was supported in part by the
National Natural Science Foundation of China under Grant 62375274, Shanghai Technology Innovation
Project under Grant XTCX-KJ-2023-01.(Corresponding authors: Tao Wu; Libing Zhou)}
\thanks{Wenbing Jiang, Yu Guo, Boyu Zhang, Xiao Chen, and Libing Zhou are with Wangzhijiang Innovation Center for Laser, Aerospace Laser Technology and System Department, Shanghai Institute of Optics and Fine Mechanics, Chinese Academy of Sciences, Shanghai 201800, China, and also with Center of Materials Science and Optoelectronics Engineering, University of Chinese Academy of Sciences, Beijing 100049, China (e-mail: lbzhou@siom.ac.cn). }
\thanks{Xuankai Xu and Lihui Jin are with the School
of Information Science and Technology, ShanghaiTech University,
Shanghai 201210, China.}
\thanks{Tao Wu is with the School
of Information Science and Technology, ShanghaiTech University,
Shanghai 201210, China, and also with the
Shanghai Engineering Research Center of Energy Efficient and Custom
AI IC, Shanghai 201210, China (e-mail: wutao@shanghaitech.edu.cn).}}

\maketitle

\begin{abstract}
The merits of thin-film surface acoustic wave (SAW) devices are pivotal to develop the high-performance intermodal acousto-optic modulators. In this work, we have proposed shear-horizontal (SH) SAW resonators for anticipated intermodal acousto-optic modulation on the thin-film lithium niobate platform. Through optimization of the cut angle of LN films, the SAW wavelength, and the thickness of interdigital transducer (IDT) electrodes, the calculated acousto-optic overlap factors utilizing SH0 modes are improved by more than an order of magnitude compared with those of Rayleigh modes. Furthermore, we have fabricated and characterized three kinds of proof-of-principle SH0 mode devices without/with grating reflectors. The electromechanical coupling coefficients ($\emph{k}_{\substack{\text{eff}}}^{\substack{2}}$) and quality factors (\emph{Q}) in the overtone resonators with grating reflectors are systematically evaluated, featuring the highest \emph{Q} of 843 with the compromised $\emph{k}_{\substack{\text{eff}}}^{\substack{2}}$ of 0.96$\%$-4.72$\%$. The results reveal that the temperature coefficients of frequency (TCF) of Rayleigh modes vary across various overtones, whereas the SH0 modes exhibit TCFs in the range of 32.3-68.9 ppm/$^\circ$C. Our fabricated SH0-mode overtone resonators demonstrate the capability of operating at power levels up to 29 dBm without electrode damage, offering a promising paradigm for robust and high-efficiency intermodal acousto-optic modulators with potential applications in integrated optical signal processing, microwave photonics, and quantum information technologies. 

\end{abstract}


\section{INTRODUCTION}
\label{sec:introduction}

Over recent decades, thin-film surface acoustic wave (SAW) devices have emerged as a highly promising candidate for next-generation high-performance RF filters, due to their notable advantages of high electromechanical coupling coefficients ($k_{\substack{\text{eff}}}^{\substack{2}}$), high quality factors (\emph{Q}), and frequency scalability compared with bulk-based devices \cite{qamar2018solidly, takai2019high, kimura2019comparative}. In particular, through transferring thin-film lithium niobate (LN) and lithium tantalate (LT) onto high acoustic velocity substrates with the Smart-Cut technology \cite{takai2019high, butaud2020innovative, zhang2020surface}, a myriad of thin-film SAW resonators and filters are extensively demonstrated \cite{he2022single, hsu2021thin, shen2021high, yang2022surface, fan2025high}. Beyond the filtering application, thin-film LN/LT SAW resonators have also been utilized for sensing and modulation \cite{hui2016plasmonic, kourani2020wideband, cai2019acousto}. For instance, leveraging the co-integration of SAW circuits and integrated optical waveguides, on-chip acousto-optic modulation (AOM) can precisely manipulate the multidimensional information of laser, such as mode, polarization, intensity, phase, and frequency, which advances the applications in optical signal processing \cite{pan2023perspective}, coherent quantum transduction \cite{shao2019microwave, sletten2019resolving}, and microwave photonics \cite{hassanien2021efficient}.

Depending on the numbers of photonics modes involved, on-chip AOM can be classified into two types: intramodal and intermodal modulation. Intramodal AOM occurs within a single photon mode, primarily for the optical phase modulation. In 2019, Cai et al. reported the first intramodal AOM fabricated on Y-cut LN-on-insulator (LNOI) by adopting the Mach-Zehnder interferometer and the high-\emph{Q} optical resonator embedded within the SAW resonator \cite{cai2019acousto}, showcasing the potential for the high-efficiency integrated AOM. Since then, a variety of approaches with different acoustic modes and integration schemes have been investigated to further optimize the modulation efficiency \cite{sarabalis2020acousto, wan2022highly, zhang2025chip}. In contrast, the intermodal AOM takes place between two distinct photon modes and enables the single-sideband amplitude modulation based on the Stokes/anti-Stokes scattering process \cite{kittlaus2021electrically}. In this sense, the intermodal AOM offers a miniaturized and integrated alternative to conventional bulk acousto-optic modulators, which typically function as frequency shifters and amplitude modulators \cite{savage2010acousto}.

So far, various acoustic modes are harnessed to implement the intermodal AOM on the thin-film LN platform. In the heterogeneous silicon and Z-cut LN platform by the transfer printing technique, non-reciprocal intermodal AOM is achieved by the Rayleigh modes with $k_{\substack{\text{eff}}}^{\substack{2}}$ of 4.82$\%$ \cite{xu2024unveiling}. The S0 mode  longitudinal leaky SAW resonator with the higher $k_{\substack{\text{eff}}}^{\substack{2}}$ on the LNOI substrate achieves the non-magnetic optical isolators based on the intermodal modulation scheme \cite{sohn2021electrically, he2025highly}. C. J. Sarabalis et al. implemented the intermodal AOM, harnessing the SH0 plate wave resonators with the resonance frequency of 440 MHz in the suspended LN films \cite{sarabalis2021acousto}. Compared with the Rayleigh and S0 modes, the SH0 mode features the higher $k_{\substack{\text{eff}}}^{\substack{2}}$ and lower energy leakage \cite{kochhar2020x}. In contrast with the SH0 plate wave resonators, SH0 mode resonators possess mechanical robustness and high power handling \cite{hsu2020large}, deemed suitable for high-performance intermodal AOM. However, comprehensive investigations of SH0 mode resonators tailored for on-chip intermodal AOM remain scarcely explored. In addition, the integration of an optical waveguide within the SAW resonator region inevitably extends the cavity length, which gives rise to the overtone response. Therefore, optimizing the merits of overtone SH0 mode resonators is crucial for intermodal AOM applications.

To enable the future integration of intermodal AOM, we have proposed and demonstrated the acoustic merits of lithium niobate-based SH0 mode resonators. For optimizing the $k_{\substack{\text{eff}}}^{\substack{2}}$ and intermodal modulation efficiency, we have comprehensively investigated the effects of various device parameters, including the cut angle of LN films, the SAW wavelength, the electrode thickness, and device structures. Furthermore, the proof-of-principle SH0 mode devices are fabricated and characterized. The overtone responses of Rayleigh and SH0 modes in our fabricated resonator are systematically investigated. The \emph{Q}-factors of overtone modes are substantially improved compared with those of the fundamental mode, reaching a moderate \emph{Q} of 843. Meanwhile, the power handling and TCF characteristics of various Rayleigh and SH0 overtones are discussed in detail.

\section{DEVICE STRUCTURE AND SIMULATION }

\subsection{Device configuration}

\begin{figure}[!t]
\centerline{\includegraphics[width=\columnwidth]{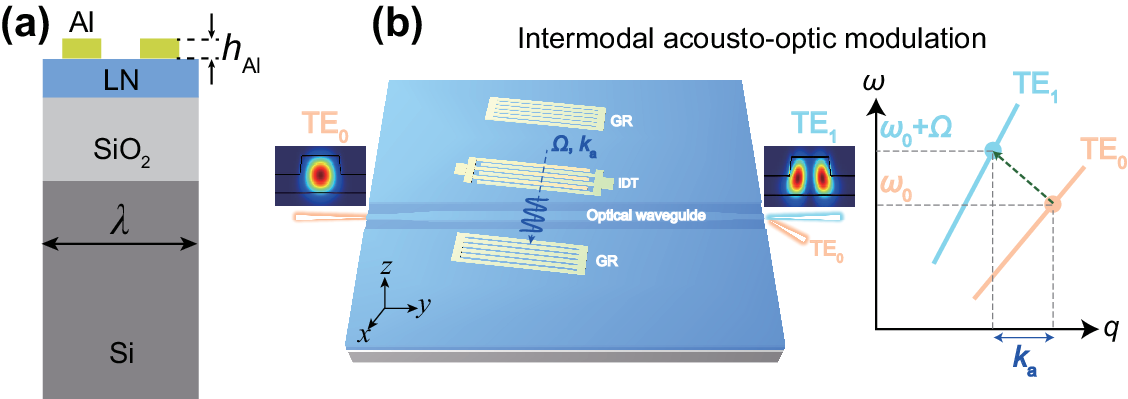}}
\caption{(a) Mock-up view of the SH0 mode resonator based on the lithium niobate-on-insulator substrate. (b) Schematic diagram of intermodal AOM based on the SAW resonator. The input TE$_0$ optical mode is modulated into the first-order optical mode TE$_1$ under interactions of the actuated SAW phonon field with the resonance frequency $\Omega/2\pi$ and the propagation wavevector $k\textsubscript{a}$. The phase-matching condition of intermodal AOM is depicted in the right panel as well.}
\label{fig1}
\end{figure}

Fig. 1(a) depicts the cross-sectional view of the SH0 mode resonator in the LNOI platform. Aluminum (Al) is selected as the electrodes of interdigital transducers (IDTs) because of its low material damping and high electrical conductivity \cite{hsu2021thin}. Fig. 1(b) shows the device configuration and schemes of intermodal AOM which primarily comprises a multimode optical waveguide and the SH0 mode resonator. The SH0 mode resonator consists of IDTs activating the SH-SAW mode and optional grating reflectors (GRs) which could confine the acoustic mode and hence improve the acousto-optic modulation efficiency. In contrast with the intramodal AOM, intermodal AOM scatters the light wave between TE$_0$ and TE$_1$ modes in our case, as shown in Fig. 1(b). With the tailored design of SAW devices, when the axial component $k\textsubscript{a}$ of propagating SAW wavevectors satisfies the momentum conservation relation, the input TE$_0$ mode could be scattered into the TE$_1$ mode \cite{kittlaus2021electrically, he2025highly}.
\begin{equation}\label{(1)}
k\textsubscript{TE1} = k\textsubscript{TE0}+k\textsubscript{a},
\end{equation}
where $k\textsubscript{TE0}$ and $k\textsubscript{TE1}$denote the wavevector of the optical TE$_0$ and TE$_1$ modes, respectively. Simultaneously, the input TE$_0$ mode with the frequency $\omega_{0}$ is scattered into the modulated TE$_1$ mode with the frequency $\omega_{0} + \Omega/2\pi$, where $\Omega/2\pi$ denotes the acoustic resonance frequency. In the light of this optical intermodal scattering, the single-sideband amplitude modulation with residual unmodulated TE$_0$ mode is realized.   

To optimize the intermodal modulation performance, two key merits should be carefully considered. The first one is $k_{\substack{\text{eff}}}^{\substack{2}}$, which determines the RF modulation bandwidth and power consumption. The second is the acousto-optic modulation efficiency, which is normally quantified by the acousto-optic overlap factor $\mathit{\Gamma}\textsubscript{ao}$. $\mathit{\Gamma}\textsubscript{ao}$ represents the spatial interaction of optical electric fields and acoustic fields, mainly stemming from two contributions of the photoelastic and cascading electro-optic effects. Due to the TE polarization we adopted, the calculation formula can be simplified as shown below

\begin{equation}\label{(1)}
\mathit{\Gamma}\textsubscript{ao} = \frac{(\sum_{i=1}^{6}p_{1i}S_{i}+\sum_{j=1}^{3}r_{1j}\varepsilon_{j})E_{1x}^{} E_{2x}^{*}}{\sqrt{\iint_A|E_{1x}|^2 dx dz \iint_A|E_{2x}|^2 dx dz}},
\end{equation}

where $p_{1i}S_{i}$ represents the contribution from the photoelastic effect, and $r_{1j}\varepsilon_{j}$ means the cascading electro-optic effect \cite{shao2019microwave,kittlaus2021electrically}. $p_{1i}$ and $r_{1j}$ are the primary photoelastic coefficients and the primary electro-optic coefficients of rotated LN, respectively \cite{weis1985lithium}. $S_{1}$, $S_{2}$, $S_{3}$, $S_{4}$, $S_{5}$, and $S_{6}$ obtained from the piezoelectric simulations (COMSOL) denote the strain tensors $S_{xx}$, $S_{yy}$, $S_{zz}$, $2S_{yz}$, $2S_{xz}$, and $2S_{xy}$ respectively. $\varepsilon_{1}$, $\varepsilon_{2}$, and $\varepsilon_{3}$ are the electric fields $\varepsilon_{x}$, $\varepsilon_{y}$, and $\varepsilon_{z}$ of the acoustic mode due to the piezoelectric effect, respectively. $E_{1x}$ and $E_{2x}$, which are extracted from the optical simulations with the Mode solver (Lumerical), describe the $x$-component electric fields of TE$_0$ and TE$_1$ modes, respectively. \emph{A} is the integration domain of the transverse LN waveguide with the normal sidewall angle of 72$^\circ$ \cite{jiang2025high}.

In the following, we will investigate the effect of the device parameters on $k_{\substack{\text{eff}}}^{\substack{2}}$ and $\mathit{\Gamma}\textsubscript{ao}$, focusing on the cut angle of LN films, the SAW wavelength, and the IDT electrode thickness.

\subsection{Effects of the cut angle and the SAW wavelength}

The cut angle of LN critically influences both $k_{\substack{\text{eff}}}^{\substack{2}}$ and the photoelastic coefficients $p_{ij}$ which strongly affect the resulting $\mathit{\Gamma}\textsubscript{ao}$. In this work, we employ $\theta$-rotated Y-X LNOI platforms because of the activated SH0 mode with the high $k_{\substack{\text{eff}}}^{\substack{2}}$ and low-loss characteristics in previous reports \cite{su2021wideband, wu2022high, lee2024spectrum, liu2025spurious}. Fig. 2(a) plots the transformed photoelastic coefficients $p\textsubscript{1j}$ of LN crystals as a function of the rotation angle $\theta$. Since the transformed $p\textsubscript{15}$ and $p\textsubscript{16}$ are zero and the strain tensor $S\textsubscript{2}$ is negligible for SH-SAWs in Y-cut LNOI, only $p\textsubscript{11}$, $p\textsubscript{13}$, and $p\textsubscript{14}$ are shown based on Eq. (2). For the SH0 mode in Y-cut LNOI, the dominant strain component is $S\textsubscript{4}$, while $S\textsubscript{1}$ and $S\textsubscript{3}$ are out of phase. Therefore, to maximize $\mathit{\Gamma}\textsubscript{ao}$, we select the cut angles that simultaneously maximize the absolute value of $p\textsubscript{14}$ and the difference between $p\textsubscript{11}$ and $p\textsubscript{13}$. As indicated in Fig. 2(a), three angles satisfy this criterion: $\theta$ = 0$^\circ$, 77$^\circ$, and 167$^\circ$. To select the optimal cut angle from three candidates, we further analyze $k_{\substack{\text{eff}}}^{\substack{2}}$ as a function of the cut angle $\theta$. Fig. 2(b) shows the simulated admittance spectra for various cut angles with the thickness of thin-film lithium niobate $h\textsubscript{LN}$ = 500 nm. As $\theta$ varies from 0$^\circ$ to 180$^\circ$, the resonance frequency of SH0 modes increases and subsequently decreases. We further calculated $k_{\substack{\text{eff}}}^{\substack{2}}$ using the standard definition
\begin{equation}\label{(2)}
k_{\substack{\text{eff}}}^{\substack{2}}=\frac{\pi^2}{8} \cdot \left(\frac{f\textsubscript{p}^2-f\textsubscript{s}^2}{f\textsubscript{s}^2}\right),
\end{equation}
where $f\textsubscript{s}$ and $f\textsubscript{p}$ denote the resonance and antiresonance frequencies, respectively\cite{kimura2019comparative, naik1998electromechanical, lu2019accurate}. As shown in Fig. 2(c), with increasing $\theta$, $k_{\substack{\text{eff}}}^{\substack{2}}$ decreases and approaches to zero at approximately $\theta$ = 130$^\circ$. For higher angles above 130$^\circ$, $k_{\substack{\text{eff}}}^{\substack{2}}$ increases rapidly. Accordingly, we select $\theta$ = 0$^\circ$ for our devices due to its higher $k_{\substack{\text{eff}}}^{\substack{2}}$ for the SH0 mode. 

After selecting the optimal cut angle, we proceed to determine the SAW wavelength $\lambda$ for efficient intermodal AOM. The modulation efficiency is strongly dependent on the relationship between $\lambda$ and the width $W\textsubscript{g}$ of the optical multimode waveguide, normally reaching a maximum at the condition of $\lambda$ = $W\textsubscript{g}$ \cite{kittlaus2021electrically, xu2024unveiling}. Therefore, the selection of $\lambda$ requires prior definition of $W\textsubscript{g}$. Given the available wafer specifications, the thicknesses of the LN film and the buried SiO$_2$ layer are fixed as 500 nm and 2 $\mu$m, respectively. Fig. 2(d) presents the simulated modal behaviors of LN rib waveguides, with optical propagation along the Z-axis to match the transverse SAW propagation. To reflect actual LNOI fabrication constraints, the sidewall etching angle is set to 72$^{\circ}$, and half-etched rib waveguides are employed to mitigate the acoustic scattering loss from the etched grooves. To enable intermodal AOM, the multimode waveguide should simultaneously accommodate TE$_\textsubscript{0}$ and TE$_\textsubscript{1}$ modes while suppressing higher-order modes to avoid the intermodal crosstalk and excess loss. As shown in Fig. 2(d), the waveguide width $W\textsubscript{g}$ of 1.7 $\mu$m meets this requirement, effectively cutting off the propagation of the higher-order TE$_2$ mode. Accordingly, the optimal SAW wavelength $\lambda$ is determined as 1.7 $\mu$m. 

\begin{figure}[!t]
\centerline{\includegraphics[width=\columnwidth]{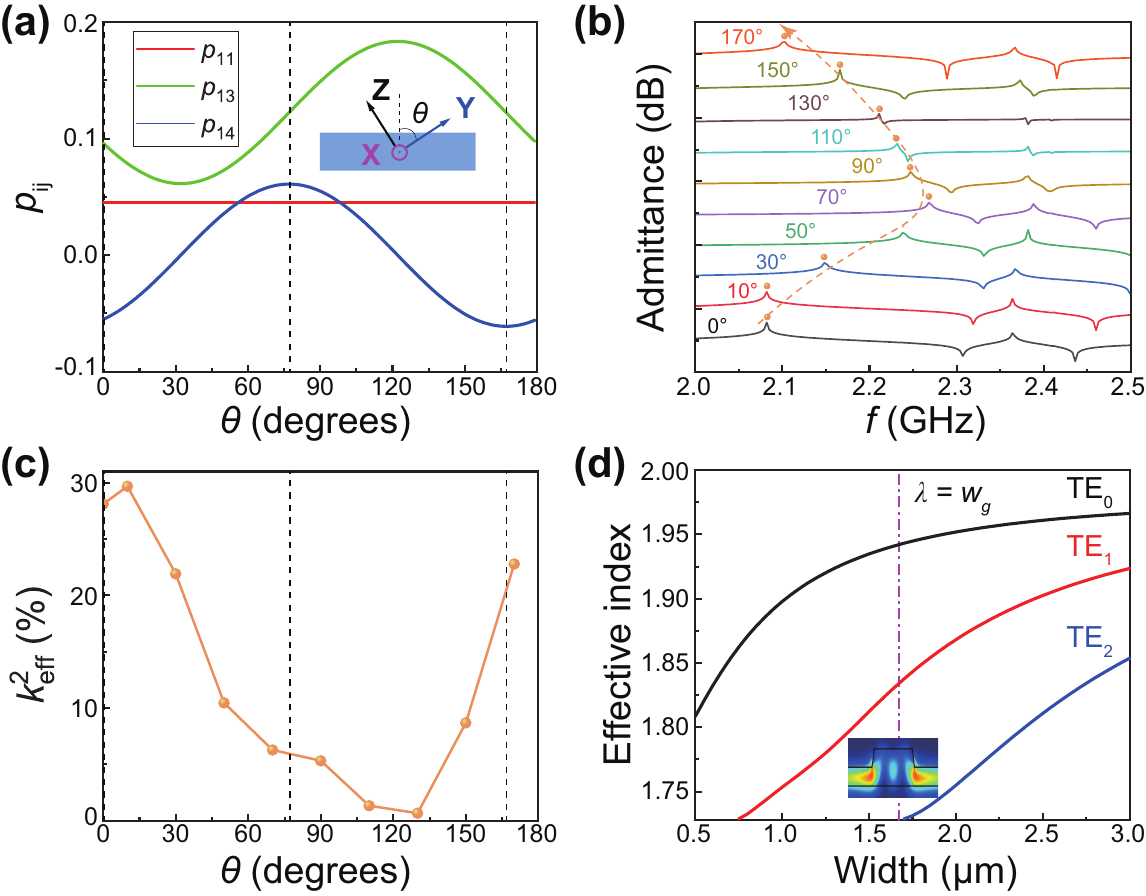}}
\caption{ (a) Transformed photoelastic coefficients $p\textsubscript{11}$, $p\textsubscript{13}$, and $p\textsubscript{14}$ as a function of the cut angle $\theta$ with respect to the Y axis. The SAW propagation direction is along the X axis. (b) Simulated admittance curves for various cut angles about the Y crystal axis. The dashed lines traces the evolution trend of SH0 modes. (c) Extracted electromechanical coupling coefficients $k_{\substack{\text{eff}}}^{\substack{2}}$ versus the cut angle $\theta$. (d) Simulated optical eigenmodes in the LN waveguide. The E-field profiles of TE$_2$ mode is displayed at the waveguide width of 1.7$\mu$m.}
\label{fig2}
\end{figure}

\subsection{Effects of the IDT electrode thickness}

Tuning the thickness of IDT electrodes is a typical approach to eliminate the spurious modes and optimize $k_{\substack{\text{eff}}}^{\substack{2}}$ of SH0 mode resonators \cite{su2021wideband,huang2023linbo3}. Meanwhile, altering IDT thickness can be harnessed to optimize the acoustic mode profiles and resulting $\mathit{\Gamma}\textsubscript{ao}$. In this study, we tune the thickness of Al electrodes $h\textsubscript{Al}$ from 100 to 450 nm with the increment of 50 nm. Utilizing finite-element-method (FEM) analyses in 2.5D piezoelectric modules of COMSOL, we have simulated the frequency dependence of admittance. To reduce the acoustic reflection from the Si substrate, a perfectly matched layer (PML) is attached to the bottom of the model. The single-cell simulation with periodic boundary conditions is adopted to reduce the simulation time. As shown in Fig. 3(a), Rayleigh modes gradually shift to lower frequency for thicker electrodes, although a shift toward higher frequency for the Rayleigh mode is observed at $h\textsubscript{Al}$ = 300 nm due to the mode coupling effect from the adjacent SH0 mode. On the other hand, the resonance frequency of the SH0 mode is remarkably decreased to about 1.2 GHz at $h\textsubscript{Al}$ = 450 nm. We note that in the intermediate thickness range of Al electrodes, the Rayleigh mode intersects the SH0 mode, deteriorating the \emph{Q}-factors of SH0 modes. The electrode thickness dependence of $k_{\substack{\text{eff}}}^{\substack{2}}$ for Rayleigh and SH0 modes is shown in Fig. 3(b), respectively. It is evident that SH0 modes attain the significantly higher $k_{\substack{\text{eff}}}^{\substack{2}}$, surpassing Rayleigh modes in terms of microwave-to-acoustic transduction efficiency and associated RF bandwidth of AOMs. The simulated $k_{\substack{\text{eff}}}^{\substack{2}}$ of the SH0 mode increases with increasing $h\textsubscript{Al}$, exhibiting a peak value of 36.1$\%$ at $h\textsubscript{Al}$ = 200 nm. However, if further increasing $h\textsubscript{Al}$ to 450 nm, $k_{\substack{\text{eff}}}^{\substack{2}}$ will decrease because of the electrode mass loading effect and the acoustic absorption of the electrodes. For Rayleigh modes, the simulated $k_{\substack{\text{eff}}}^{\substack{2}}$ is 4.73$\%$ at $h\textsubscript{Al}$ =100 nm, and subsequently increases due to the SH0 mode hybridization effect as $h\textsubscript{Al}$ approaches 250 nm. Similar to SH0 modes, the $k_{\substack{\text{eff}}}^{\substack{2}}$ of Rayleigh modes decreases for thicker electrodes.

Furthermore, we have calculated the $\mathit{\Gamma}\textsubscript{ao}$ for both Rayleigh and SH0 modes with various $h\textsubscript{Al}$ using eq.(2), in order to determine the optimal electrode thickness. Fig. 4(a) and (b) show the simulated strain tensors $S_{1}$, $S_{3}$, and $S_{4}$ for the Rayleigh and SH0 modes at $h\textsubscript{Al}$ = 100 nm, respectively. For the Rayleigh mode, $S_{1}$ is the dominant component, while the shear strain tensor $S_{4}$ is negligible. Meanwhile, $S_{1}$ and $S_{3}$ reverse their signs along the $y$-direction, which reduces $\mathit{\Gamma}\textsubscript{ao}$ and leads to weak modulation efficiency. Nevertheless, for the SH0 mode, the shear strain tensor $S_{4}$ dominates over $S_{1}$ and $S_{3}$. More importantly, all strain components preserve the same sign along the vertical $y$-direction. As a result, the antisymmetric strain profiles along the \emph{x}-direction overlap constructively with the antisymmetric TE$_1$ optical mode profiles shown in Fig. 1(b), yielding a substantial enhancement of $\mathit{\Gamma}\textsubscript{ao}$. 

The corresponding calculated spatial distributions of $\mathit{\Gamma}\textsubscript{ao}$ for Rayleigh and SH0 modes are illustrated in Fig. 4(c) and (d), respectively. For the Rayleigh mode, the overlap distribution oscillates along the $x$-axis and changes sign along the $y$-direction, resulting in significant cancellation of $\mathit{\Gamma}\textsubscript{ao}$ upon integration. In contrast, the SH0 mode preserves a uniform sign of $\mathit{\Gamma}\textsubscript{ao}$ across the waveguide region. In addition, the magnitude of $\mathit{\Gamma}\textsubscript{ao}$ for the SH0 mode is notably larger than that of the Rayleigh mode, as evident from Fig. 4(c) and (d), indicating substantially stronger acousto-optic modulation efficiency. Fig. 4(e) plots the integral $\mathit{\Gamma}\textsubscript{ao}$, including contributions from both the photoelastic and cascading electro-optic effects, as a function of $h\textsubscript{Al}$ for both modes. For the SH0 mode, $\mathit{\Gamma}\textsubscript{ao}$ initially increases with $h\textsubscript{Al}$, reaches a maximum at $h\textsubscript{Al}$ = 150 nm, and then declines toward a plateau. For comparison, $\mathit{\Gamma}\textsubscript{ao}$ for the Rayleigh mode exhibits vastly smaller values due to the cancellation effect of its strain profiles. Its value remains relatively unchanged with increasing $h\textsubscript{Al}$ until an abrupt jump occurs at $h\textsubscript{Al}$ = 250 nm, which we attribute to the more concentrated strain distribution arising from intermodal coupling with the adjacent SH0 mode. 

Based on the above analysis, we define $h\textsubscript{Al}$ as 150 nm to achieve the maximal $\mathit{\Gamma}\textsubscript{ao}$ while maintaining a relatively high $k_{\substack{\text{eff}}}^{\substack{2}}$. At the optimized $h\textsubscript{Al}$, the calculated $\mathit{\Gamma}\textsubscript{ao}$ for the SH0 mode exceeds that of the Rayleigh modes by more than an order of magnitude. For the acoustic resonator design, we set the IDT aperture to 40$\lambda$ to ensure the long acousto-optic coupling length while mitigating acoustic diffraction loss from IDTs. The electrode number of IDTs $N\textsubscript{t}$ is optimized to achieve the impedance matching, minimizing the power reflection of IDTs.

\begin{figure}[!t]
\centerline{\includegraphics[width=\columnwidth]{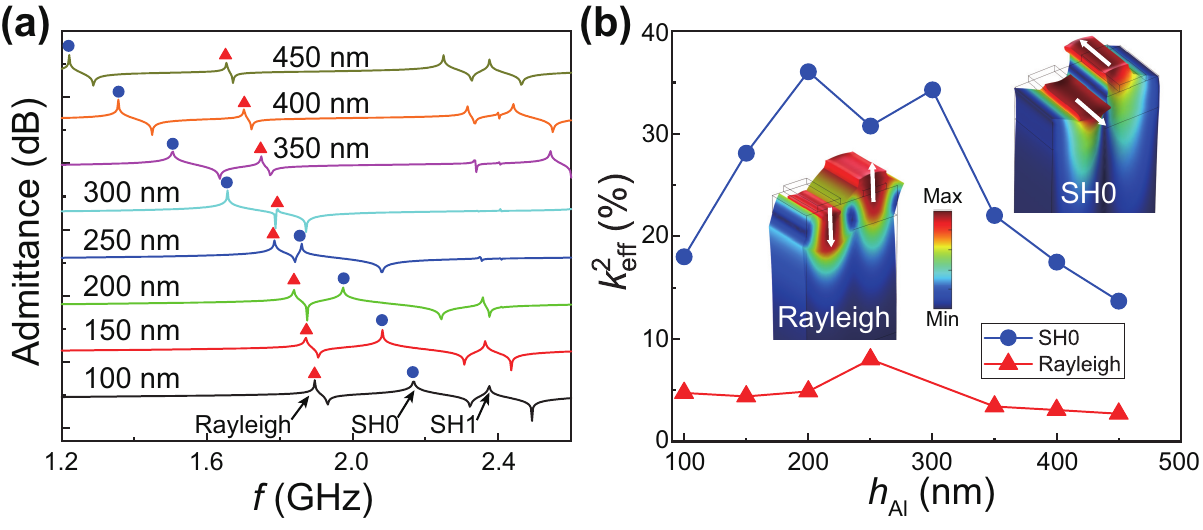}}
\caption{ (a) Simulated admittance spectra with tuning the Al electrode thickness ranging from 100 to 450 nm at the increment of 50 nm. (b) Extracted $k_{\substack{\text{eff}}}^{\substack{2}}$ of the Rayleigh and SH0 modes, respectively. The displacement profiles of Rayleigh and SH0 modes are shown in the inset, respectively.}
\label{fig3}
\end{figure}

\begin{figure}[!t]
\centerline{\includegraphics[width=\columnwidth]{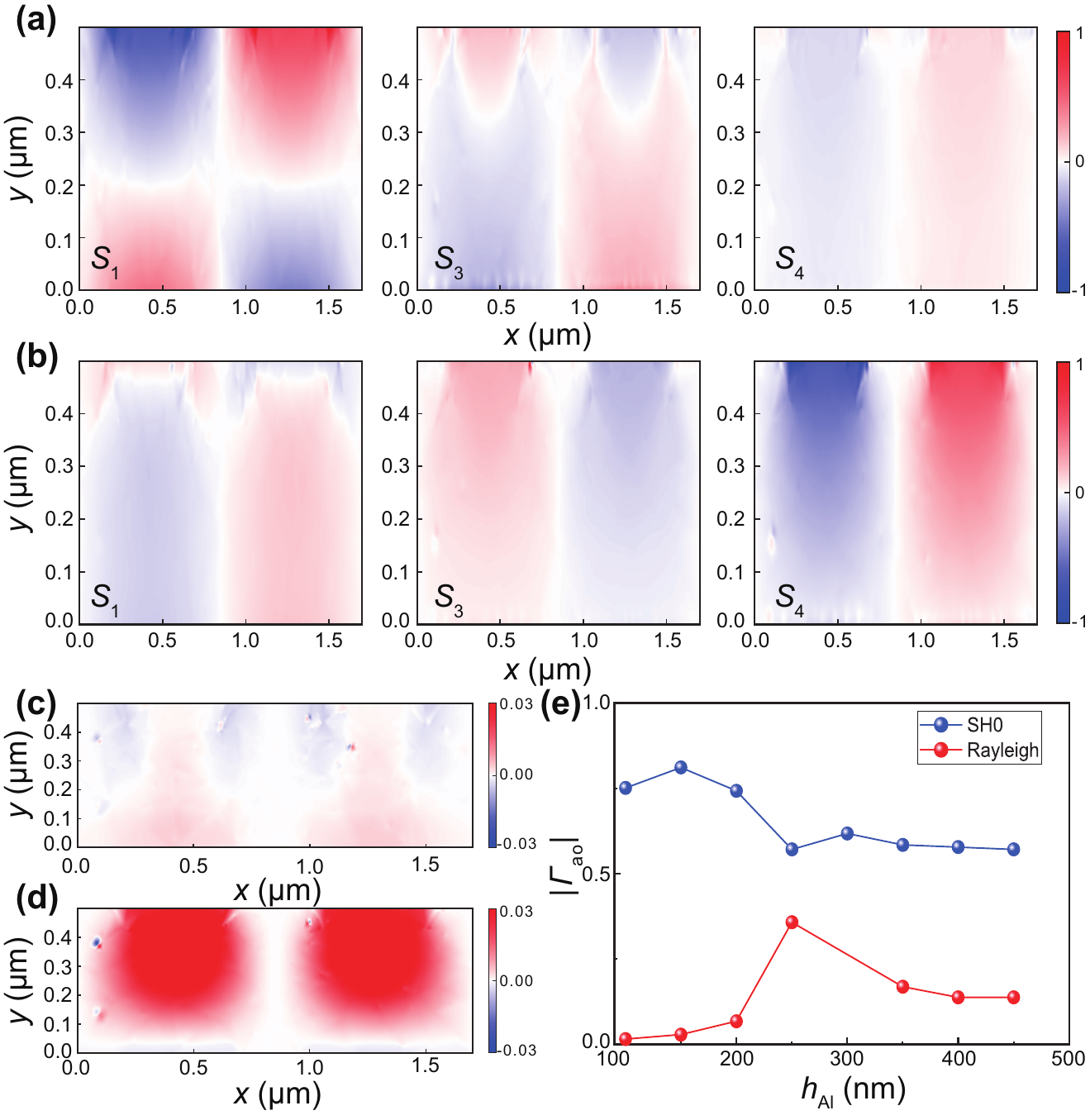}}
\caption{ Acoustic strain tensors of $S_{1}$, $S_{3}$, and $S_{4}$ of the Rayleigh mode (a) and the SH0 mode (b) with $h\textsubscript{Al}$ = 100 nm, respectively. Calculated spatial distributions of the AO overlap distribution in the optical waveguide for the Rayleigh mode (c) and the SH0 mode (d) with $h\textsubscript{Al}$ = 100 nm. (e) Comparison of $\mathit{\Gamma}\textsubscript{ao}$ based on Rayleigh and SH0 modes with respect to the Al-electrode thickness $h\textsubscript{Al}$.}
\label{fig4}
\end{figure}

\section{DEVICE FABRICATION AND MEASURED RESULTS}

\subsection{Device fabrication}

The fabrication of our SAW devices was started with a Y-cut LN/SiO$_2$/Si substrate. We adopted two-step photolithography and lift-off preocesses to fabricate the designed devices. In the first-step photolithography, electron beam lithography (EBL) was undertaken to define the sub-micron pattern structures of IDTs and grating reflectors (GRs). Then, 150 nm of Al films were deposited with electron-beam evaporation, and lifted off to form the IDT electrodes and GRs. The second-step photolithography with a MA-6 mask aligner (SUSS MicroTec) was adopted to pattern padding layers. To reduce the series resistance $R\textsubscript{s}$ of the pad layer, A thicker gold (Au) padding layer with 700-nm-thickness was patterned by the electron-beam evaporation and a lift-off process. 

Fig. 5 shows the optical microscope images of three fabricated SAW devices, labelled as Device A,  Device B, and Device C. Device A has a single IDT with the electrode width $a$ of 425 nm, the pitch $p$ of 850 nm, the IDT aperture \emph{W} of 40$\lambda$, and $N\textsubscript{t}$ of 41. Device B and Device C have the same IDT structure as Device A. In particular, Device B adds a GR on the one side of the IDT, emitting the unidirectional SAW for intermodal AOM. Device C adds two symmetrically distributed GRs on both sides which form an acoustic resonator. The gap between IDTs and GRs in Device C is designed to be 50 $\mu$m, accommodating the LN multimode optical waveguide. The electrode width and the pitch of GRs are identical with those of IDTs. The GR number $N\textsubscript{g}$ is set as 50. A tilted angle of 6.6$^\circ$ of SAW propagation direction around the X axis is determined because of the required phase matching condition of intermodal AOM. We also performed the scanning electron microscopy (SEM) measurements on three kinds of devices, as shown in the zoomed-in images of Fig. 5. All the measured electrode geometry sizes in three fabricated devices are close to the design values, and the fabricated electrode fringes are well uniform and clean without any residual resists, indicating the reproducibility and reliability of the fabrication process. The detailed device geometry parameters are listed in Table I.    

\begin{figure}[!t]
\centerline{\includegraphics[width=\columnwidth]{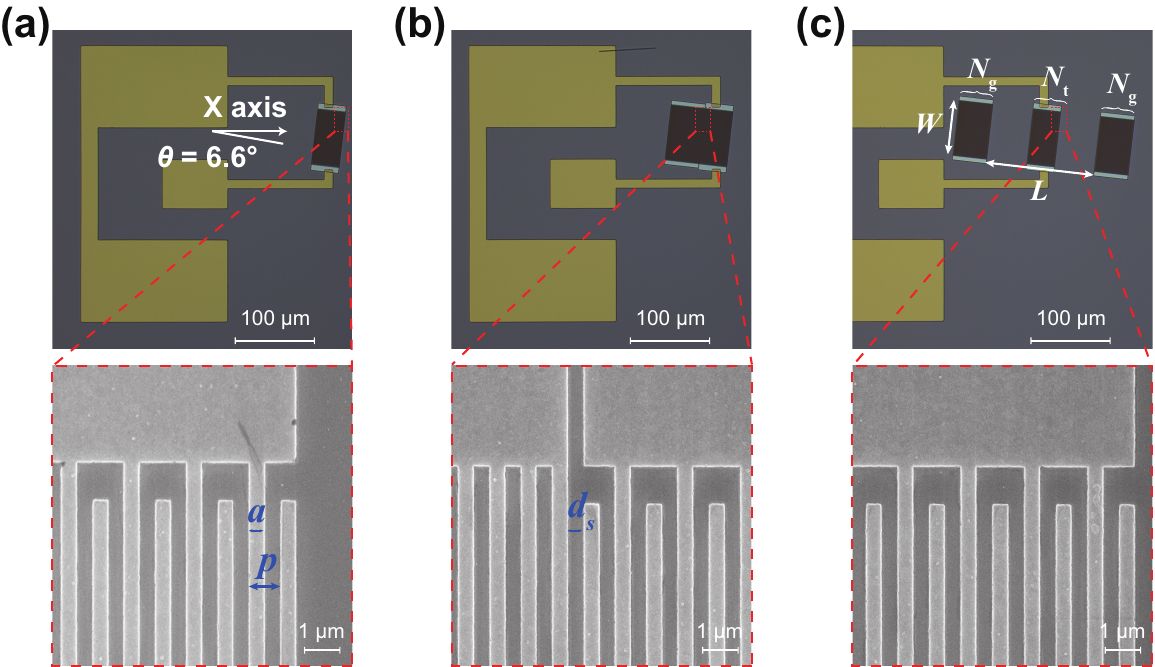}}
\caption{ Optical microscope images of the fabricated Device A (a), Device B (b), and Device C (c). The zoomed-in regions are the corresponding SEM images in each SAW resonator, respectively.}
\label{fig5}
\end{figure}

\begin{table}
\caption{Key parameters of three kinds of devices}
\centering
\label{table}
\renewcommand{\arraystretch}{1.5}
\setlength{\tabcolsep}{7pt}
\begin{tabular}{l c c c c c r}
\hline
\hline
\textbf{Device} & \textbf{A} & \textbf{B} & \textbf{C} \\
\hline
Electrode width (\emph{a}) & 425 nm & 425 nm & 425 nm \\ 
Pitch width (\emph{p}) & 850 nm & 850 nm & 850 nm \\
IDT electrode number ($N\textsubscript{T}$) & 41 & 41 & 41 \\
Grating number ($N\textsubscript{g}$) & 50 & 50 & 50 \\
IDT aperture (\emph{W}) & 68 $\mu$m & 68 $\mu$m & 68 $\mu$m \\
Gap between IDT and GR ($d\textsubscript{s}$) & / & 0.425 $\mu$m & 50 $\mu$m \\
\hline
\hline
\end{tabular}
\label{tab1}
\end{table}

\subsection{Multimode resonator feature}

\begin{figure}[!t]
\centerline{\includegraphics[width=\columnwidth]{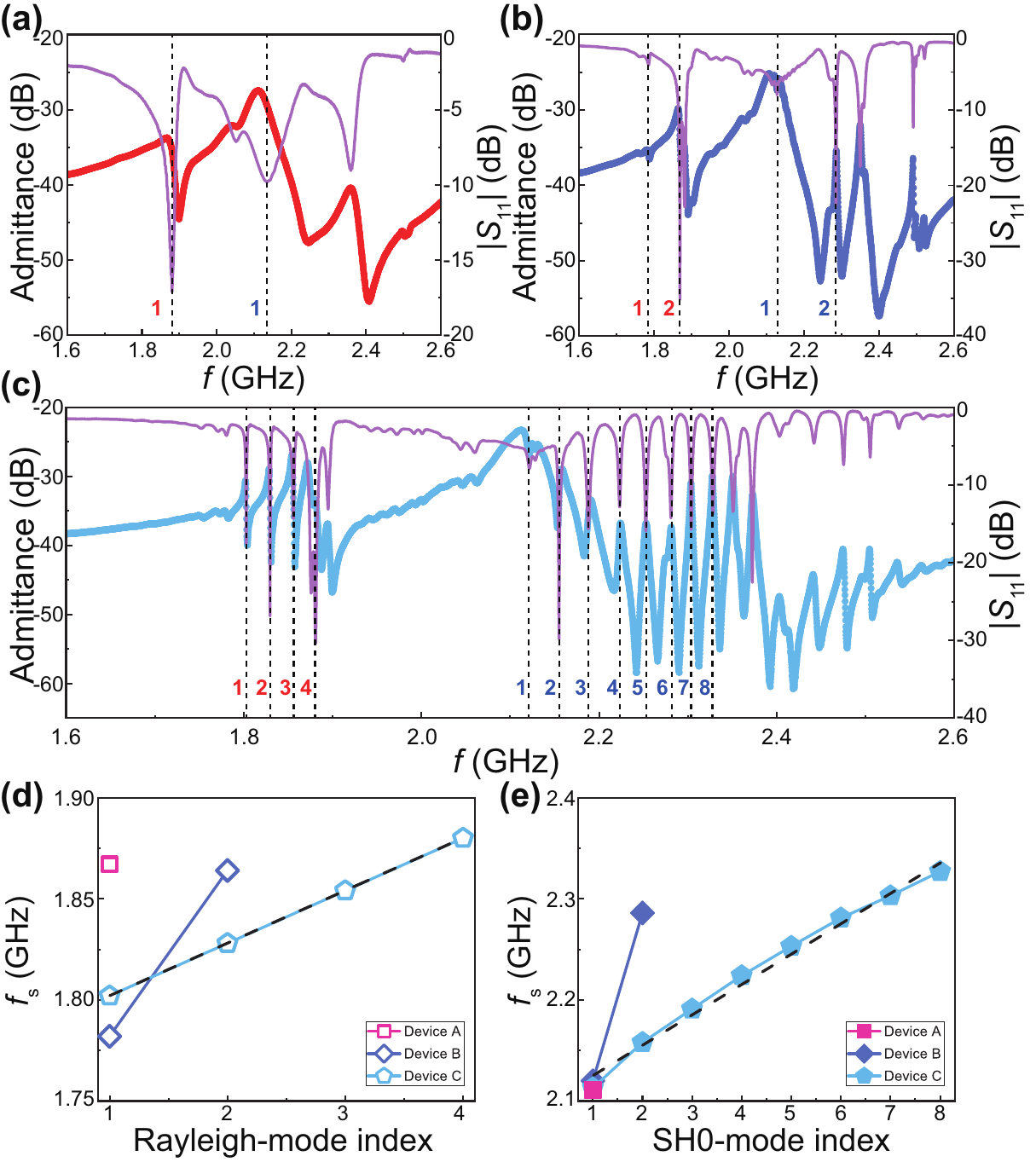}}
\caption{ (a) Measured admittance and reflection coefficient of Device A (a), Device B (b), and Device C (c). The Rayleigh and SH0 mode index are indicated in red and blue numbers, respectively. (d) Summarized Rayleigh-mode index and the corresponding resonance frequency $f\textsubscript{s}$ in all three devices. The dashed line denotes the linear fit with a coefficient of determination ($R^2$) of 1. (e) Summarized SH0-mode index and the corresponding resonance frequency $f\textsubscript{s}$ in all three devices. The dashed line denotes the linear fit with $R^2$ of 0.989.}
\label{fig6}
\end{figure}

Characterization of the fabricated devices was performed in ambient dry air with a vector network analyzer (Keysight PNA-L N5234B) equipped with ground-signal-ground probes. The admittance $Y\textsubscript{11}$ of each device was measured and recorded after the calibration. The corresponding reflection coefficients $S\textsubscript{11}$ were converted from the measured admittance using the relation
\begin{equation}\label{(3)}
S\textsubscript{11} = \frac{Y_0 - Y_{11}}{Y_0 + Y_{11}}.
\end{equation}
$Y_0$ is the admittance load, typically being 1/50 $\Omega^{-1}$. Fig. 6(a)-(c) shows the magnitude of admittance $|Y|$ and reflection coefficient $|S\textsubscript{11}|$ of Device A, Device B, and Device C, respectively. For Device A, three main resonances are observed at 1.867 GHz, 2.111 GHz, and 2.359 GHz, in correspondence to the Rayleigh mode, the SH0 mode, and the higher-order SH (SH1) mode. The small discrepancy of approximately 29 MHz between the measured and simulated resonance frequency arises from fabrication deviations of IDTs and uncertainties of material parameters used in the simulations. Compared with the resonance frequencies in Fig. 3(a), the measured resonance frequencies $f\textsubscript{s}$ are well consistent with the FEM predictions. The dips of $|S\textsubscript{11}|$ in three devices also coincide with the corresponding peaks of $|Y|$, indicating the nature of acoustic resonance. For Device B and Device C, the implementation of GRs forms the acoustic resonator, exhibiting a number of overtone responses marked in Fig. 6(b) and (c). Compared with Device B, the longer cavity length of Device C results in more overtones, as shown in Fig. 6(c). Above the resonance frequency of the SH0 overtones, we also observe multiple spurious responses originating from SH1 overtones. These spurious modes can be suppressed by tuning the thickness of IDT electrodes, LN films, and the SiO$_2$ layer \cite{huang2023linbo3}, to further optimize the $k_{\substack{\text{eff}}}^{\substack{2}}$ and $Q\textsubscript{s}$ of SH0 modes. 

Fig. 6(d) and (e) summarize the resonance frequency of Rayleigh and SH0 overtones, respectively. Device A features a single Rayleigh and SH0 mode, while Device B exhibits two Rayleigh tones and two SH0 tones, and Device C exhibits four Rayleigh tones and eight SH0 tones which are equally spaced due to the constructive interference effect in the Fabry-Perot acoustic cavity \cite{lu20205, lin2024thin}. We also observed the multiple resonances above 2.35 GHz in device C, which correspond to SH1 overtones which are not marked in Fig. 6(c). The average Rayleigh overtone spacings of Device B and Device C which can be defined as the phonon free spectral range (FSR) are 82 (84) MHz and 26 (24) MHz for $f\textsubscript{s}$ ($f\textsubscript{p}$), respectively. Accordingly, the average SH0-mode FSR of Device B and Device C are 151 (58) MHz and 30.6 (26.1) MHz for $f\textsubscript{s}$ ($f\textsubscript{p}$), respectively. The decreased FSR in Device C is attributed to the longer cavity length, consistent with the trend in our reported thin-film SAW resonators \cite{jiang2023thin}. Compared with the constant Rayleigh-mode FSR, the SH0-mode FSR in Device C exhibits fluctuating and a decreasing trend for higher-order overtones, which is attributed to the impedance mismatch of SH0 overtones \cite{kurosu2023impedance}.

Fig. 7(a) and (b) present the admittance responses of the elaborated 3rd-order Rayleigh mode and the 7th-order SH0 mode, respectively. The overtone responses are free from spurious modes, which is beneficial for achieving high \emph{Q}-factors. The corresponding $k_{\substack{\text{eff}}}^{\substack{2}}$ and $Q\textsubscript{s}$ are calculated and compared in Fig. 7(c) and (d), respectively. Herein, we focus on $Q\textsubscript{s}$ which is defined by the 3dB-\emph{Q} at $f\textsubscript{s}$, as the \emph{Q}-factors at $f\textsubscript{s}$ directly relate to the modulation efficiency. Fig. 7(c) displays the evolution of $k_{\substack{\text{eff}}}^{\substack{2}}$ for all the overtone responses of Rayleigh and SH0 modes. For Device A (with a single IDT), the $k_{\substack{\text{eff}}}^{\substack{2}}$ of the Rayleigh mode is 4.27$\%$, consistent with the simulated value. When the SAW resonator supports overtones in Device B and Device C, $k_{\substack{\text{eff}}}^{\substack{2}}$ gradually decreases to below 1$\%$ in Device C. Meanwhile, the fundamental SH0 mode exhibits a large $k_{\substack{\text{eff}}}^{\substack{2}}$ of 16.4$\%$ in Device A, and decreases to 12.9$\%$ in Device B, and further decreases to the range of 0.96$\%$-4.72$\%$ in Device C. Notably, lower-order SH0 mode reveals higher $k_{\substack{\text{eff}}}^{\substack{2}}$, manifesting their more prone excitations. 

In contrast with the evolution of $k_{\substack{\text{eff}}}^{\substack{2}}$, the extracted $Q\textsubscript{s}$ of Rayleigh and SH0 overtones have substantially enhanced compared with those in Device A. For instance, the Rayleigh-mode $Q\textsubscript{s}$ increases from 12 in Device A to the range of 316-418 in Device C, with the 3rd-order Rayleigh mode achieving the maximum $Q\textsubscript{s}$ of 418. Similarly, the SH0 mode $Q\textsubscript{s}$ rises from 26 in Device A to the range of 62-843 in Device C. Contrary to the trend observed in $k_{\substack{\text{eff}}}^{\substack{2}}$, higher-order overtones exhibit significantly higher $Q\textsubscript{s}$ for both Rayleigh and SH0 modes, as shown in Fig. 7(d). Although Device C achieves a moderate $Q\textsubscript{s}$ of 843 for the 7th-order SH0 mode, there remains room for further improvement through optimization of the LN film thickness and suppression of spurious modes \cite{zhang2023ultra, li2022high}. 

Based on the coupled mode theory, and assuming equal optical loss for the TE$_0$ and TE$_1$ modes \cite{zhou2024nonreciprocal}, the intermodal AOM efficiency $\eta^2$ can be expressed as
\begin{equation}\label{(3)}
\eta^2 = \sin^2\left(\frac{gQ\textsubscript{s}}{\pi f\textsubscript{s}c}\sqrt{\frac{\emph{P}\textsubscript{i}(1-|S_{11}|^2)n\textsubscript{g1}n\textsubscript{g2}Wt}{\hbar Q\textsubscript{c}}}\right),
\end{equation}
where $\emph{P}\textsubscript{i}$, $|S_{11}|$, and $Q\textsubscript{c}$ represent the driving RF power, the magnitude of the reflection coefficient at $f\textsubscript{s}$, and the coupling \emph{Q}-factor of acoustic modes to external feedline circuits, respectively. \emph{g}, \emph{t}, and \emph{W} denote the acousto-optic coupling strength calculated by $\mathit{\Gamma}\textsubscript{ao}$ \cite{shao2019microwave}, the acoustic energy ratio of the LN layer extracted from the piezoelectric simulations, and the IDT aperture, respectively. \emph{n}\textsubscript{g1} and \emph{n}\textsubscript{g2} are the group index of TE$_0$ and TE$_1$ modes, which can be obtained from the optical simulations. We note that the $|S\textsubscript{11}|$ at all resonances is below -8 dB, with the minimum reaching -30 dB, indicating that the SH0 mode resonators operate near the critical-coupling regime for which $Q\textsubscript{c}$ = 2$Q\textsubscript{s}$. The calculated intermodal AOM efficiency for four modes as a function of the applied RF power is shown in Fig. 8(a). For the fundamental Rayleigh mode in Device A and the 3rd-order Rayleigh mode in Device C, $\eta^2$ reaches -44.65 dB and -35.17 dB at $\emph{P}\textsubscript{i}$ = 30 dBm which represents a 1 dB extrapolation beyond the demonstrated power range shown below, respectively. Because the $\mathit{\Gamma}\textsubscript{ao}$ of the SH0 mode has been improved by over an order of magnitude than that of the Rayleigh mode, the corresponding $\eta^2$ is significantly enhanced. According to our analyses, at $\emph{P}\textsubscript{i}$ = 30 dBm, $\eta^2$ reaches -16.07 dB for the fundamental SH0 mode in Device A and -11.2 dB for the 7th-order SH0 mode in Device C, respectively. Fig. 8(b) also displays the calculated $\eta^2$ for various SH0 overtones with the RF driving power of 30 dBm. The trend of $\eta^2$ coincides with that of $Q\textsubscript{s}$ across all the SH0 overtones. The observed largest $\eta^2$ for the 7th-order SH0 mode in Device C results from the stronger acousto-optic coupling strength of SH0 mode and the maximal $Q\textsubscript{s}$ of 843 in the overtone SH0 mode resonators.
 
\begin{figure}[!t]
\centerline{\includegraphics[width=\columnwidth]{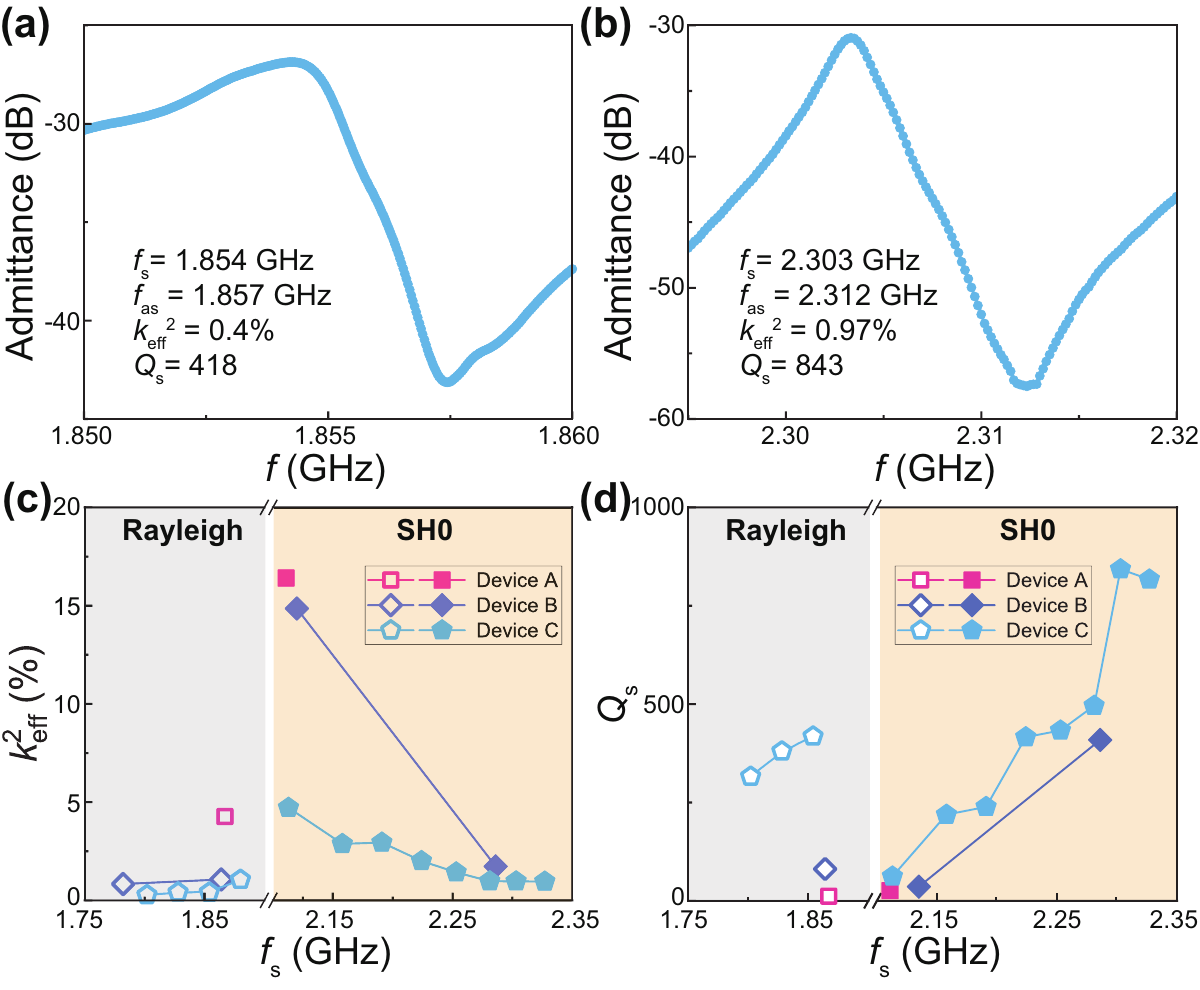}}
\caption{ (a) Measured admittance of the 3rd-order Rayleigh mode. (b) Measured admittance of the 7th-order SH0 mode. Extracted $k_{\substack{\text{eff}}}^{\substack{2}}$ (c) and $Q\textsubscript{s}$ (d) of Rayleigh and SH0 modes of the three devices.}
\label{fig7}
\end{figure}

\begin{figure}[!t]
\centerline{\includegraphics[width=\columnwidth]{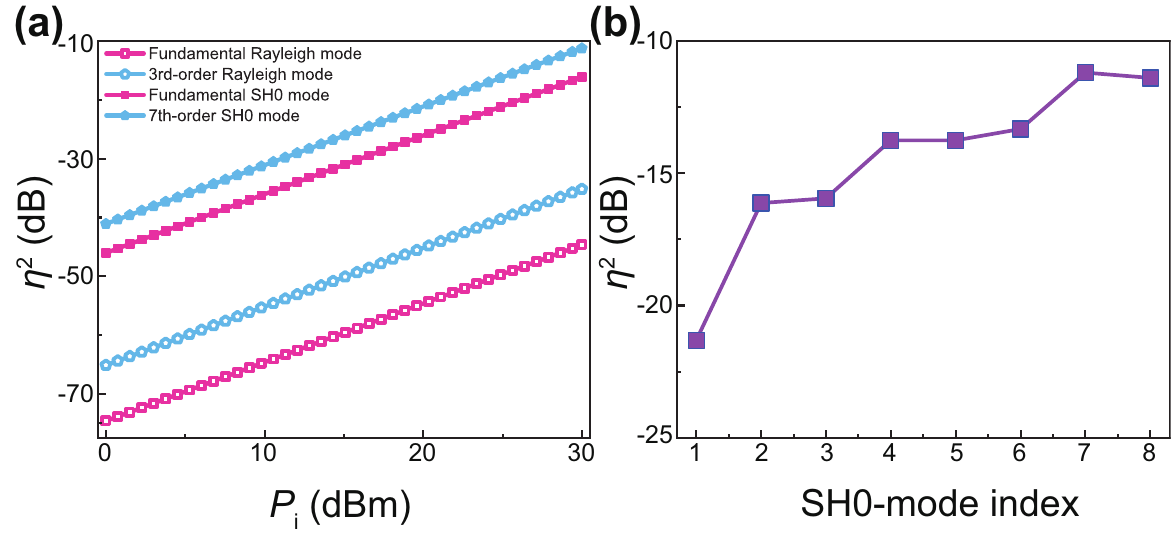}}
\caption{ (a) Calculated AOM modulation efficiency $\eta^2$ with respect to the driving RF power based on the measured parameters of acoustic resonators. (b) Evolution of $\eta^2$ for various SH0 overtones with the RF driving power of 30 dBm.}
\label{fig8}
\end{figure}

\subsection{Power handling}

\begin{figure}[!t]
\centerline{\includegraphics[width=\columnwidth]{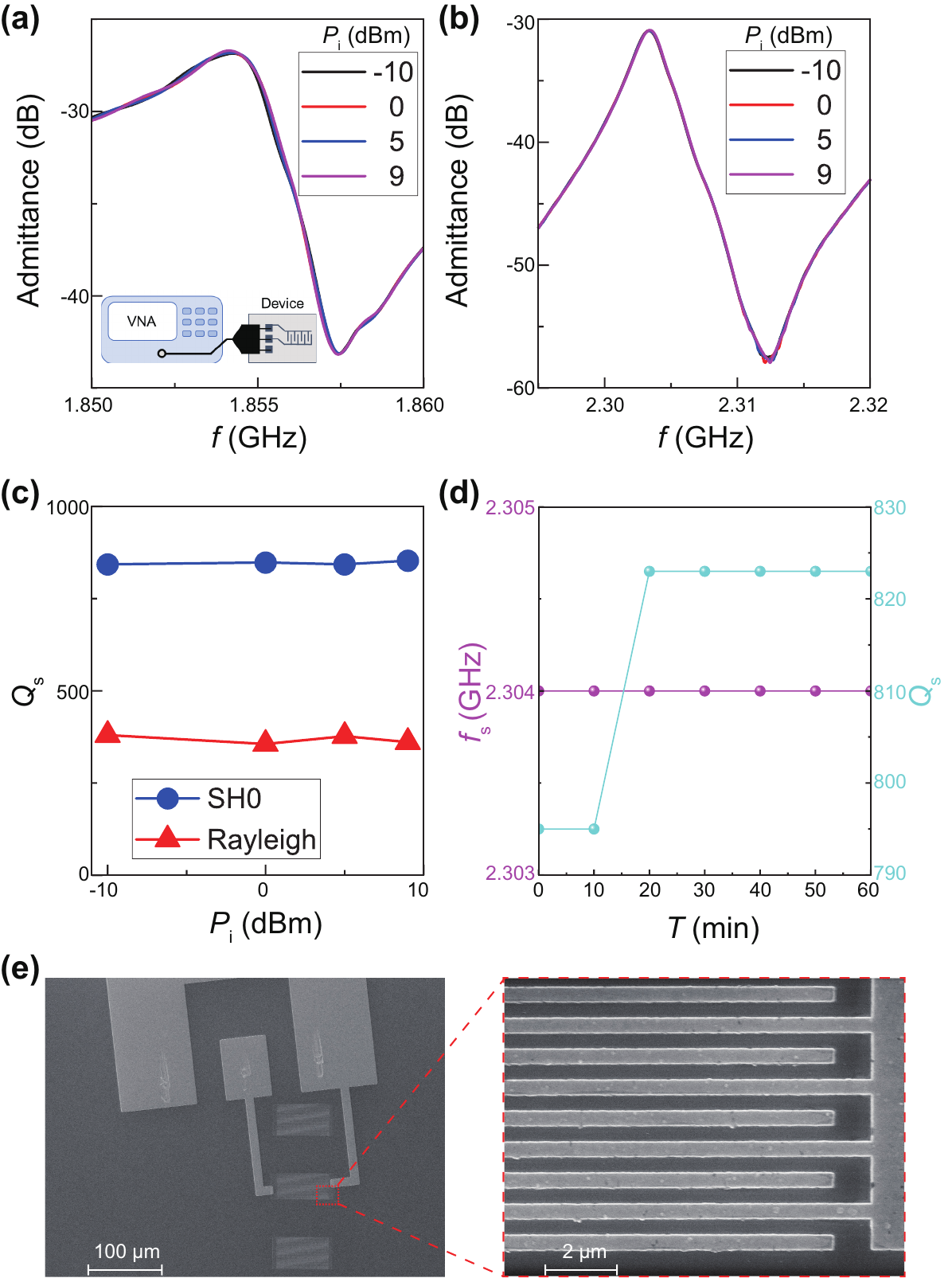}}
\caption{ Admittance response of the 3rd-order Rayleigh mode (a) and the 7th-order SH0 mode (b) of device C with the applied power increasing from -10 to 9 dBm. The inset illustrates the schematic of power-handling measurement setups. (c) Measured $Q\textsubscript{s}$ of both modes as a function of applied power. (d) Extracted resonance frequency $f\textsubscript{s}$ and $Q\textsubscript{s}$ for the 7th-order SH0 mode in one-hour duration measurements. (e) Magnified SEM images of Device C after high-power RF driving of 29 dBm.}
\label{fig9}
\end{figure}

As inferred from Eq.(5), the intermodal AOM efficiency $\eta^2$ is proportional to $\sqrt{\emph{P}\textsubscript{i}}$. Therefore, achieving high-power operation of SAW resonators is essential for enhancing the modulation efficiency. Additionally, the power handling capability of SAW resonators is also crucial for high-performance RF filter applications \cite{shen2021high, hsu2024c}. To evaluate the power robustness, we have measured the admittance of three devices under increasing the RF power from -10 to 9 dBm which is the upper limit of the used vector network analyzer, as shown in the inset of Fig. 9(a). All the Rayleigh and SH0 overtones exhibit strong stability against increasing RF power. Fig. 9(a) and (b) show two elaborated resonances of device C for the 3rd-order Rayleigh mode and the 7th-order SH0 mode, respectively. Over the entire power range, the magnitude of admittance $|Y|$ at $f\textsubscript{s}$ varies by less than 0.2 dB, while the extracted $f\textsubscript{s}$ and $Q\textsubscript{s}$ of both modes remain almost unchanged, as summarized in Fig. 9(c). 

The long-term reliability under fixed power operations is also critical for the practical AOM application. We have conducted one-hour continuous measurements, during which the admittance data were recorded every 10 minutes under the maximal input power of 9 dBm. The measured admittance remains almost unchanged across all sweeps, indicating well long-term spectral stability of our devices. The corresponding $f\textsubscript{s}$ and $Q\textsubscript{s}$ for the 7th-order SH0 mode over one hour are displayed in Fig. 9(d), where $f\textsubscript{s}$ remains very stable without any changes and $Q\textsubscript{s}$ varies within 2.45$\%$ in one hour. It is observed that the measured $Q\textsubscript{s}$ increases slightly after 10 minutes, presumably arising from the self-heating effect during the continuous power measurements, which expels the moisture absorbed from the air within the device \cite{sui2025miniaturized,fang2025hybrid,qian2026suppressing,shen2022saw,gonzalez2021method}. Due to the reflection coefficients $|S\textsubscript{11}|$ for the 3rd-order Rayleigh mode and the 7th-order SH0 mode are below -14 dB, as shown in Fig. 6(c), the reflected partition of input power due to impedance mismatch can be negligible in the power handling measurements.  

To assess the power handling under higher-power conditions for intermodal AOM, we have applied the high-power RF signals in continuous wave (CW) mode to Device C using a power amplifier capable of delivering RF power up to 31 dBm, and examined the devices after the CW RF excitations using the SEM measurements. Fig. 9(e) shows the magnified SEM images taken after 10 minutes of RF driving test at $f\textsubscript{s}$ = 2.303 GHz, with an actual power of 29 dBm at the probe tips due to 2 dB loss from RF cables and connectors. RF driving tests at other $f\textsubscript{s}$ have been conducted as well. As a result, no electrode damage is clearly observed in Fig. 9(e), indicating that the devices can operate at high power levels of at least 29 dBm constrained by our measurement setup. These results confirm that the fabricated SAW resonators can reliably operate at high power levels up to 29 dBm for high-power operations, making them suitable for implementation in high-efficiency acousto-optic modulators.

\subsection{Temperature coefficient of frequency}

\begin{figure}[!t]
\centerline{\includegraphics[width=\columnwidth]{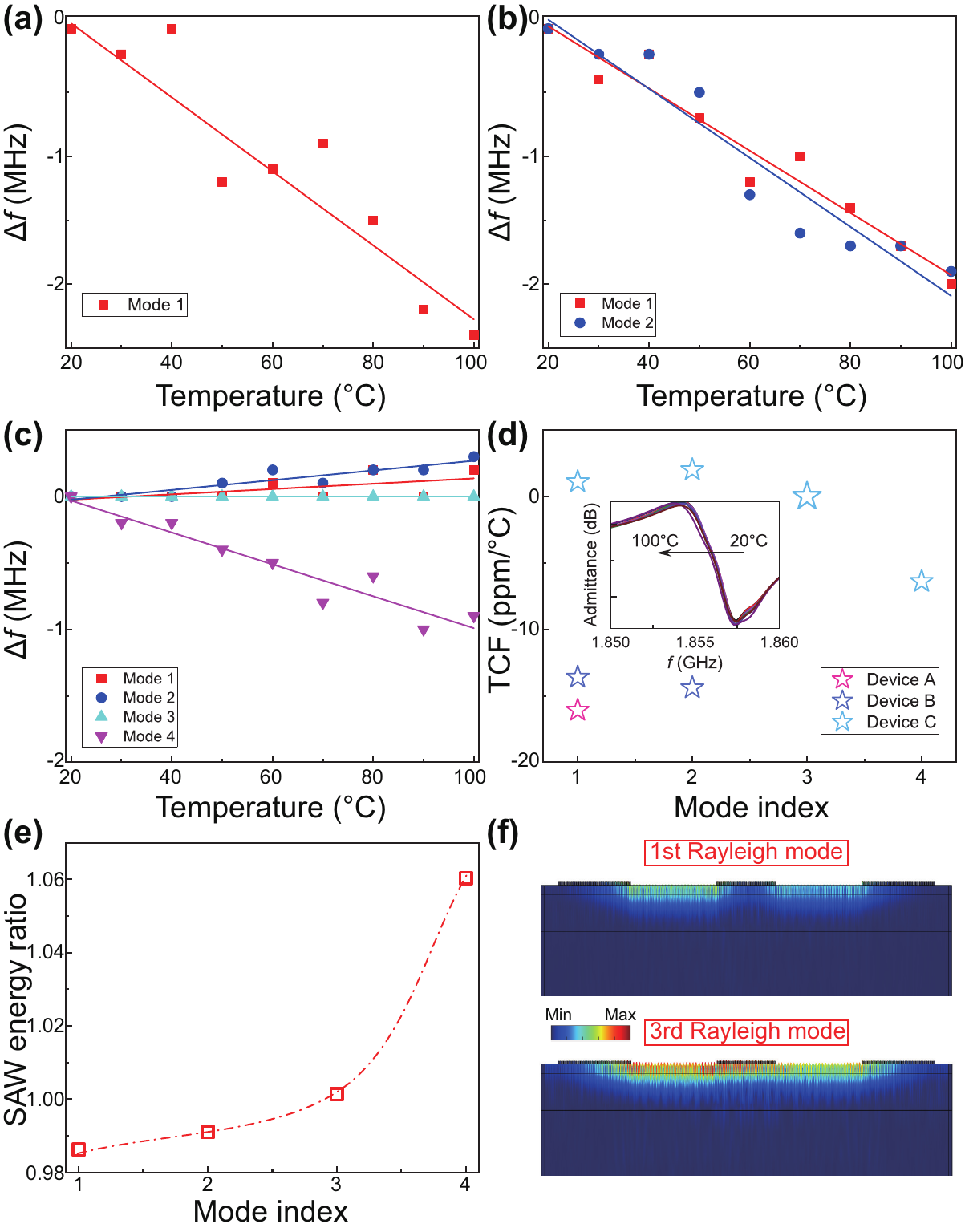}}
\caption{ Temperature dependence of anti-resonance frequency $f\textsubscript{p}$ of Rayleigh overtones in device A (a), device B (b), and device C (c). The solid lines are linear fitting of the experimental data. (d) Extracted TCF values of various Rayleigh modes in three kinds of devices. The inset plots the measured admittance spectra of the 3rd-order Rayleigh mode in device C, with temperature increasing from 20 to 100 $^\circ$C. (e) Simulated SAW energy ratio in the LN layer to that in the SiO$_2$ layer for various Rayleigh modes. (f) Simulated displacement profiles of the 1st-order and 3rd-order Rayleigh modes, respectively.}
\label{fig10}
\end{figure}

\begin{figure}[!t]
\centerline{\includegraphics[width=\columnwidth]{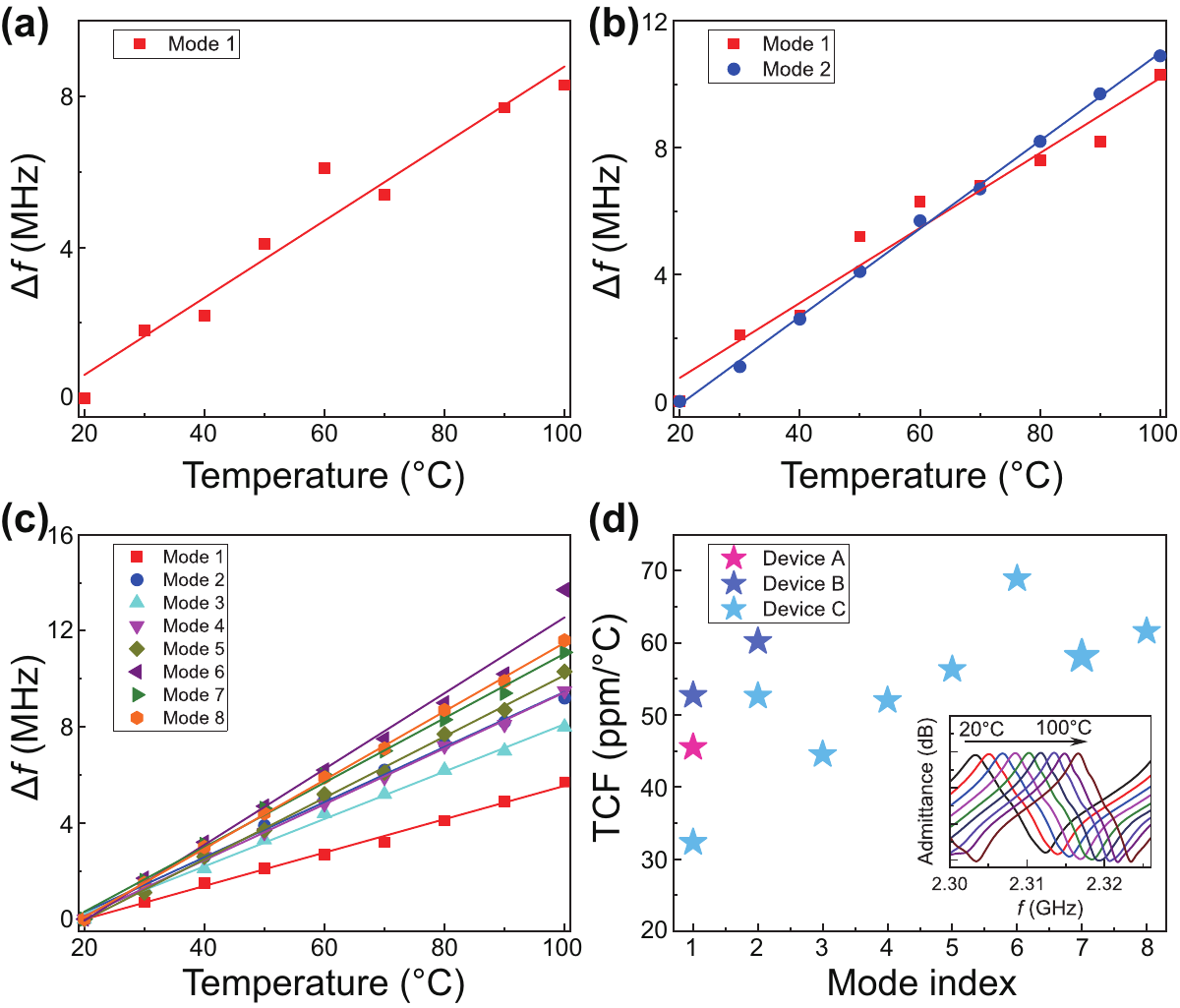}}
\caption{Temperature dependence of anti-resonance frequency $f\textsubscript{p}$ of SH0 overtones in device A (a), device B (b), and device C (c). The solid lines represent the corresponding linear fittings. (d) Extracted TCF values of various SH0 modes in three kinds of devices. The inset plots the measured admittance spectra of the 7th-order SH0 mode in device C, with temperature increasing from 20 to 100 $^\circ$C.}
\label{fig11}
\end{figure}

We have further analyzed the temperature coefficients of frequency (TCFs) of all the overtone modes for three devices. Since there are spurious modes near $f\textsubscript{s}$ of the fundamental SH0 mode, in the following discussion we have analyzed the TCF of $f\textsubscript{p}$, which is compared with and verified by the TCF values extracted from $f\textsubscript{s}$. Fig. 10(a)-(c) show the measured antiresonance frequency shift of Rayleigh modes in Device A,  Device B, and Device C in the temperature range of 20-100$^\circ$C, respectively. For Device A and Device B, $f\textsubscript{p}$ shifts to lower frequency with increasing temperature. The average TCFs for these two devices are -16.1 ppm/$^\circ$C and -14 ppm/$^\circ$C, respectively. For Device C, the TCF values of the 1st-order and 2nd-order Rayleigh modes are positive, while the TCF values of the 3rd-order and 4th-order Rayleigh modes are zero and negative, respectively. The zero-TCF characteristic of the 3rd-order Rayleigh mode, which is verified by the TCF simulations as well, can be seen in the inset of Fig. 10(c). The variation of TCFs among various Rayleigh overtones stems from their distinct spatial SAW energy distribution. We have calculated the SAW energy ratio, which is defined as the ratio of the SAW energy distributed in the LN layer to that in the SiO$_2$ layer, as displayed in Fig. 10(e). For lower-order Rayleigh overtones, the SAW energy is distributed equally in the LN layer with a negative TCF and the SiO$_2$ layer with a positive TCF \cite{liu2025tuning}. As a result, the device TCF is compensated, reaching zero-TCF for the 3rd-order Rayleigh mode. However, for higher-order Rayleigh overtones, the distributed SAW energy ratio in the LN layer increases, leading to the negative TCF observed. 

Fig. 11(a)-(c) display the measured frequency shift of SH0 modes in Device A, Device B, and Device C, respectively. Fig. 11(d) summarizes the TCF values for SH0 overtones in three devices. In contrast with Rayleigh modes, the $f\textsubscript{p}$ of SH0 overtones reveal a linear positive shift upon increasing temperature, as illustrated in the inset of Fig. 11(d). The corresponding TCF values for the SH0 overtones range from 32.3 to 68.9 ppm/$^\circ$C, which is comparable to the reported TCFs of SH0 mode resonators \cite{liu2025tuning, Li2019temperature, liu2023spurious}. However, for the SH0 fundamental mode, the TCF values extracted from $f\textsubscript{s}$ and $f\textsubscript{p}$ are -32.4 ppm/$^\circ$C and 45.5 ppm/$^\circ$C, respectively, comparable to the corresponding simulated TCF values of -65 ppm/$^\circ$C and 59.7 ppm/$^\circ$C. The opposite TCF behaviors extracted from $f\textsubscript{s}$ and $f\textsubscript{p}$ for the SH0 fundamental mode arise from the different SAW energy concentration ratio in the LN layer at $f\textsubscript{s}$ and $f\textsubscript{p}$. At the higher frequency $f\textsubscript{p}$, the SAW energy penetrates more into the SiO$_2$ layer, hence resulting in positive TCFs. Similarly, for SH0 overtones with higher resonance frequencies, because the SAW energy ratio in the SiO$_2$ layer gradually increases according to TCF simulations, both TCF values extracted from $f\textsubscript{s}$ and $f\textsubscript{p}$ become positive in the range of 32.3 to 68.9 ppm/$^\circ$C, consistent with the trend previously observed in the LNOI platform \cite{hsu2021wideband}. The comparison between the overtone SH0-mode resonators in this work and those reported in the literature is shown in Table II \cite{kourani2020wideband, lin2024thin, Li2019temperature, lu2018rf}. Our fabricated devices achieve the highest resonance frequency, and the $Q\textsubscript{s}$ surpasses the devices on the unsuspended LNOI platform. Although the overtone SH0 mode resonators on the suspended LNOI platform exhibit higher $Q\textsubscript{s}$, the devices may suffer from the power handling capabilities and nonlinear effects under high-power conditions.

\begin{table*}[t]
\caption{Comparison of overtone SH0-mode resonators}
\centering
\label{table}
\renewcommand{\arraystretch}{1.5}
\setlength{\tabcolsep}{7pt}
\begin{tabular}{c c c c c c}
\hline
\hline
Ref. & Structure & $f\textsubscript{s}$ (GHz) & $Q\textsubscript{s}$ & $k_{\substack{\text{eff}}}^{\substack{2}}$ & TCF (ppm/$^\circ$C) \\
\hline
$[12]$ & Suspended X-10$^\circ$Y-LiNbO$_3$ & 0.3 -- 0.5 & 1650  -- 3000 & 0.59\% -- 2.56\% & / \\ 
$[47]$ & Suspended 36$^\circ$YX-LiNbO$_3$/SiO$_2$/Si & 0.084 -- 0.515 & 450 -- 3000 & 1.5\% -- 12\% & 8.96 -- 14.7 \\ 
$[49]$ & Suspended X-10$^\circ$Y-LiNbO$_3$ & 0.1 -- 0.683 & 1043 -- 3633 & 0.13\% -- 2.03\% & / \\ 
$[39]$ & X-10$^\circ$Y-LiNbO$_3$/SiO$_2$/Si & 0.536 -- 0.62 & 345 -- 476 & 1.27\% -- 3.67\% & -42.0 -- -38.5 \\ 
This work & YX-LiNbO$_3$/SiO$_2$/Si & 2.113 -- 2.327 & 62 -- 843 & 0.96\% -- 4.72\% & 32.3 -- 68.9 \\ 
\hline
\hline
\end{tabular}
\label{tab2}
\end{table*}

\section{CONCLUSION}

In this work, we have designed and experimentally demonstrated the tailored SH0 mode resonator in the Y-cut LNOI platform for the future applications of intermodal acousto-optic modulation. The SAW wavelength and the IDT electrode thickness are systematically optimized to facilitate high $k_{\substack{\text{eff}}}^{\substack{2}}$ and $\mathit{\Gamma}\textsubscript{ao}$. 
Simulation results demonstrate that the strain profiles of SH0 modes, maintaining the uniform sign along the normal direction, constructively overlap with optical modes for intermodal AOM which leads to higher modulation efficiency than Rayleigh modes. The calculated $\mathit{\Gamma}\textsubscript{ao}$ of SH0 modes is improved by over an order of magnitude compared with that of Rayleigh modes. 

Based on the optimized device parameters, we have fabricated and characterized three kinds of SAW devices: without GRs, with one GR, and with two GRs. Compared with the device without GRs, the resonators incorporating one or two GRs support multiple overtones and exhibit the enhanced $Q\textsubscript{s}$ for SH0 modes, which is helpful for improving modulation efficiency, attaining a maximal $Q\textsubscript{s}$ of 843. The overtone SH0-mode resonators also demonstrate the capability of operating at power levels up to 29 dBm without electrode damage, serving as a promising acoustic platform for high-efficiency intermodal acousto-optic modulators. Furthermore, the TCF for overtone SH0-modes is positive, ranging from 32.3 ppm/$^\circ$C to 68.9 ppm/$^\circ$C, whereas the TCF of overtone Rayleigh-modes varies from positive to negative, or even to zero due to the different SAW energy distribution in the LN and SiO$_2$ layers. Our results have established a feasible avenue for high-performance intermodal acousto-optic modulators, and also offered a potential candidate for wide-tuning range oscillators and quantum acoustics applications.

\bibliography{Reference}
\bibliographystyle{IEEEtran}

\end{document}